# Capacity Planning for Cluster Tools in the Semiconductor Industry

Martin Romauch[a,*], Richard F. Hartl[a]

[a] *Department of Business Administration*
*University of Vienna, Oskar-Morgenstern-Platz 1, 1090 Wien*

---

**Abstract**

This paper proposes a new model for Cluster-tools with two load locks. Cluster-tools are widely used to automate single wafer processing in semiconductor industry. The load locks are the entry points into the vacuum of the Cluster-tool's mainframe. Usually there are two of them available. Each lot being processed, is dedicated to a single load-lock. Therefore at most two different lots (with possibly different processing times and qualification) can be processed simultaneously. This restriction is one of the major potential bottlenecks.

Capacity planning is one of the possible applications for the proposed model and the paper demonstrates the integration into a more general framework that considers different tool types and different operational modes.

The paper also generalizes an earlier model that is limited to three processing chambers. The proposed modeling approach is based on makespan reductions by parallel processing. It turns out that the performance of the new approach is similar, when compared to the generalized model for three chambers, but the new approach outperforms the generalized model for four and more chambers.

---

## 1. Introduction

This paper considers Cluster-Tools, a tool type that can be found in semiconductor fabrication front-end sites, also called wafer fabrication facilities. The wafer is the substrate used for the fabrication of integrated circuits for semicon-ductor devices. A wafer is a disc, usually a slice of mono-crystalline silicon with a width between 150 and 300 mm, and a thickness between 0.5 and 1 mm.

*Corresponding author
*Email addresses:* `martin.romauch@univie.ac.at` (Martin Romauch), `richard.hartl@univie.ac.at` (Richard F. Hartl)

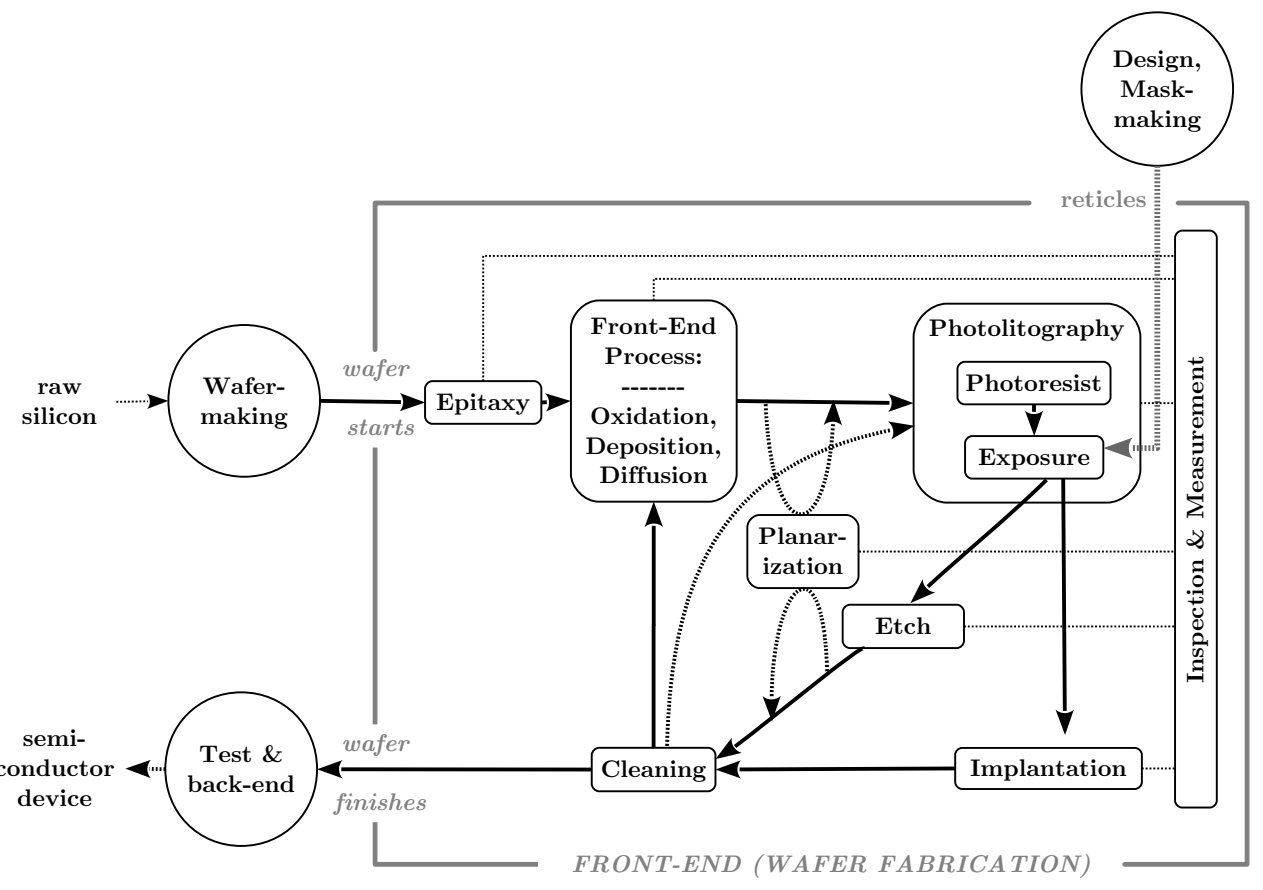

Figure 1: front end fabrication as a re-entrant flow shop (based on Sorenson [24] and Mönch et al. [19])

The input of front-end facilities are containers of wafers (also called lots or FOUPs, Front Opening Unified Pods). A lot usually consists of 20 or more wafers that require the same processing. The fabrication is characterized by a large number loops or layers (compare Hutcheson [15]). Each loop consists of one or more processing steps; one of them is a lithographic process that allows to cover parts of the wafer (photo resist) from the process that follows. Figure 1 gives a schematic view on wafer fabrication.

The flow diagram in Figure 1, is a representation of the fabrication process as a re-entrant flow shop. According to Thiesse and Fleisch [25] the traversed path of a single lot can be several kilometers long.

In semiconductor industry most of the equipment is automated and characterized by load/unload operations, robot handling, testing, alignment, cooling and much more. Cluster Tools (compare Franssila [6]) for instance, have one or more load locks and multiple processing chambers. Cluster-tools are used to automate several process types. The internal software defines the behavior of the system. Cluster tools allow an automatic transfer of wafers between load ports and process chambers with possibly different processes inside a vacuum. The system can also be used for parallel processing to increase throughput and productivity. The wafers are transferred between chambers under a vacuum using a robotic arm to prevent exposure to air to prevent oxidation and contamination.

Since a Cluster-tool may operate on different lots at the same time, the combination of lots is important and the Cluster-tool may show different operation cycle times for the same type of lot. According to Mönch et al. [19] several researchers studied the optimization of internal scheduling, but the corresponding models are not suitable to consider several Cluster-tools at the same time. One of the major bottlenecks of a Cluster-tool are the load locks (see Christopher [3]). The load locks are the entry point into the vacuum of the Cluster tool.

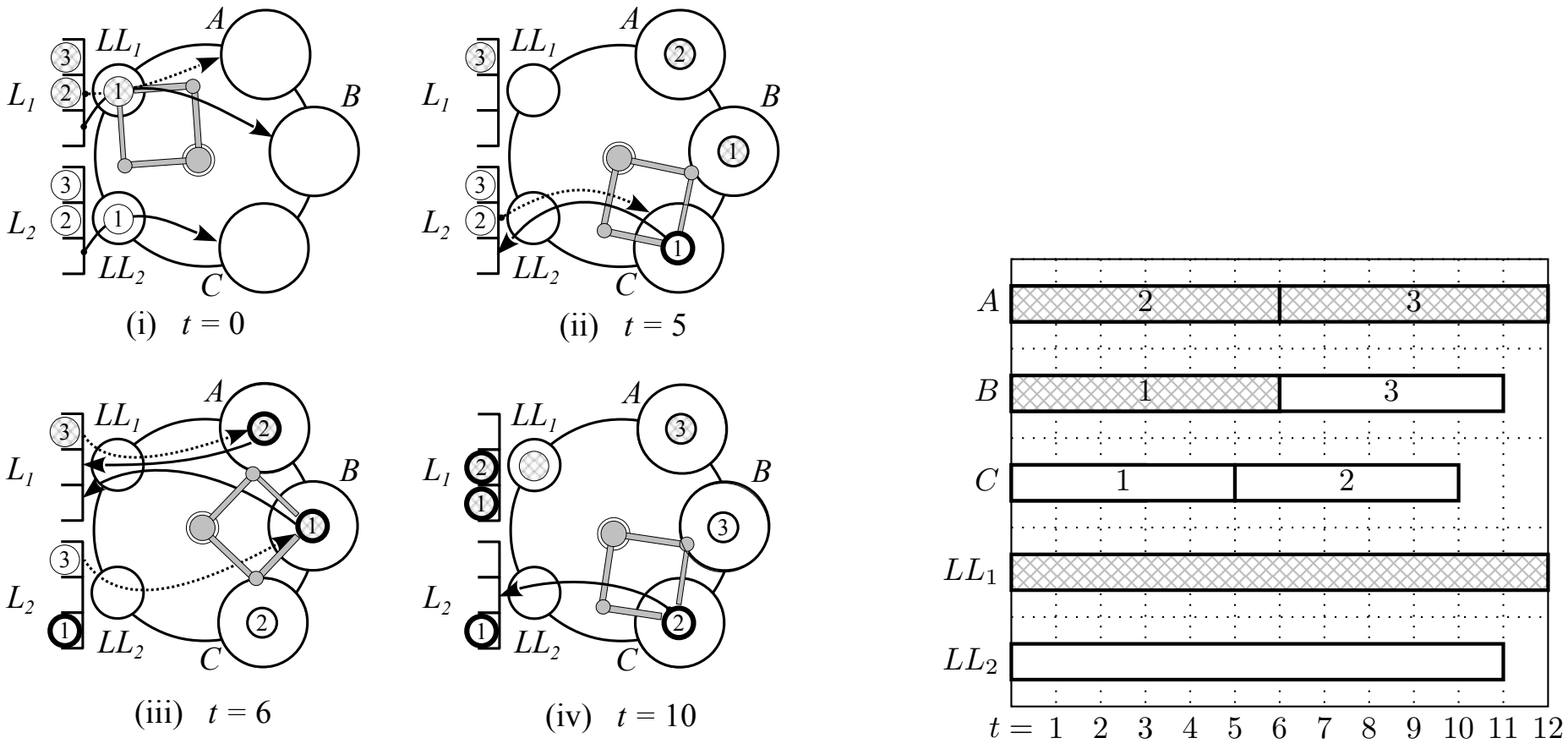

Figure 2: A Cluster-tool with two load locks that is processing wafers in parallel. On left hand side: four snapshots of the Cluster-tool for $t = 0, 5, 6, 10$ can be found. The thickness of the outline of the wafers, indicates the processing progress. The corresponding Gantt-chart can be found on the right hand side of the Figure.

In order to describe a Cluster-Tool with respect to load locks and chambers, a simplified schedule for a Cluster-tool that is working in parallel mode is discussed in Example 1. The example is based on Figure 2.

**Example 1.** On the left hand side of Figure 2, a Cluster-tool that is working in parallel mode with two load locks $LL_1$ and $LL_2$ is illustrated. In this example the chambers are supposed to be identical. Later in the example, one of the chambers will be less flexible than the others. In the beginning ($t = 0$), each load lock is occupied by one lot ($LL_1$ with lot $L_1$ and $LL_2$ with lot $L_2$), and each lot consists of three wafers. Each wafer needs to be processed in one of the chambers. The load lock $LL_1$ is occupied with the lot $L_1$ that needs the longer processing time. More precisely, each wafers from $L_1$ need a processing time of six time units. For $L_2$ the processing time is five time units per wafer. In this example the wafer handling time is neglected, therefore it is possible to start all processes in each chamber at the same time.

In the first step (i) with $t = 0$, two wafers from lot $L_1$ are assigned to chambers $A$ and $B$; and one wafer from $L_2$ is assigned to chamber $C$. At $t = 5$ the wafer in chamber $C$ is finished and it is passed back to $L_2$. Then, the second wafer from $L_2$ is assigned to chamber $C$. At $t = 6$ the wafers in $A$ and $B$ are handed back to $L_1$ and chamber $A$ will be occupied by the last wafer of $L_1$; chamber $B$ will be occupied by the last wafer of $L_2$.

The corresponding Gant-chart can be found on the right hand side of Figure 2. The make-span is 12 with an idle time percentage of $\frac{1}{12}$ on average. If wafers from $L_1$ are excluded from chamber $C$ then at least two wafers from $L_1$ need to be assigned to the same chamber ($A$ or $B$), hence the schedule is optimal with respect to the make-span.

Now, additionally suppose that the problem is scaled by thirty, and each lot counts 90 wafers. Then the wafers can be distributed in such a way that the idle time percentage vanishes. If this is possible, then the total processing time is $90 \cdot (6 + 5) = 990$. Therefore it is sufficient to find a distribution where each chamber finishes after 330 time units. Note that 330 is divisible by five and six:

$$330 = \overbrace{55 \cdot 6}^{A} = \overbrace{35 \cdot 6 + 24 \cdot 5}^{B} = \overbrace{66 \cdot 5}^{C}$$

Therefore 55 wafers from $L_1$ can be assigned to chamber $A$ and 66 wafers of $L_2$ can be assigned to chamber $C$. The remaining wafers can be assigned to chamber $B$.

This example shows that on a larger scale where larger quantities are considered, a continuous approximation - where it is allowed to computationally "split" wafers - leads to reasonable results.

LP based Cluster-tool models can be directly integrated in capacity models or in master planning like proposed in Ponsignon and Mönch [22] and Romauch and Klemmt [23]. The combination of simulation and optimization (LP) as discussed in Almeder et al. [1] Gansterer et al. [8] and Juan et al. [16] is also suitable for the integration of LP based Cluster-Tools models.

Finally, real-time dispatching that integrates LP approaches can be found in Doleschal et al. [4] and Ham et al. [11] which can be extended to LP based Cluster-tool models. According to Duemmler and Wohlleben [5] there are also various alignments necessary to achieve an effective WIP Flow Management that assures that the planned capacity consumption (utilization) is coherent with the dispatching reality. Therefore, improvements of LP based Cluster-tool models are important to several areas.

The focus of this paper is on static capacity planning, but it is important to note that the LP models introduced in this paper may also be useful to other applications, like master planning, simulation and dispatching.

In the context of capacity planning for front-end facilities, a concise prediction of the equipment utilization is one of the major tasks. For given demands - that is wafer starts for given product routes - a calculation needs to confirm that the capacity restrictions are met. Furthermore it is necessary to identify hard/soft bottlenecks that are limiting the throughput.

Depending on the purpose and time frame, there are several methods to tackle this capacity planning problem - sometimes elementary MRP calculations are sufficient, but the large majority is covered by simulation and linear programming. This paper is dedicated to linear programming formulations and proposes a Cluster Tools model for multiple chambers and two load locks that is performant, flexible and easy to analyze.

The contribution of this paper is threefold. First, we present a new model that considers more than three chambers. Second we derive the model presented in Ortner [21] in a more general way that also allows to handle than three chambers. Third, computational experiments are used to compare the performance of the proposed methods.

## 2. Capacity Planning in Semiconductor Industry

In order to show the applicability of the proposed Cluster-Tool model, the integration into a basic capacity planning model will be presented. The discussed model is static, the processing sequence is not considered and the quantities of wafers are continuous. Aspects like multiple time-frames can be understood as an elementary extension (cf. Romauch and Klemmt [23]), therefore the discussion (and the computational experiments) is restricted to a basic model to prove the concept and to avoid a cumbersome notation.

The discussion of input and output of the basic capacity planning model starts with an example that is depicted in Figure 3. The input for the capacity planning model is the *weekly going rate* $d_p$ for each product $p \in P$ on the basis of lots for a given fixed lot size. That means, $d_p$ wafers of product product $p$ need to be produced in the following week. Each product corresponds to a product route that consists of processing steps. Some of the processing steps can be considered equivalent for different product routes and these equivalent processing steps constitute the so called job classes ($j \in J$). The capacity consumption (e.g. service times) and flexibility for members of the same job class is the same. Analogous to Figure 3 the primary demand $d_p$ on basis products infers a secondary demand $\lambda_j$ on basis job classes. For instance job class 1 has a demand $\lambda_1 = 30$ because this job class occurs once in the product routes for product one and product two ($\lambda_1 = d_1 + d_2$).

The wafers $\lambda_j$ are distributed among the available machines, i.e. $y_{ji}$ wafers from job class $j$ are assigned to machine $i$. In total, for each job class $j$ the demand $\lambda_j$ needs to be satisfied, i.e. $\sum_{i \in \{a,b,c,d\}} y_{ij} = \lambda_j$. The distribution that corresponds to $y_{ji}$ infers a load on the corresponding machines that depends on the service time $t_{ji}$, i.e. $u_i = \sum_{j=1}^{6} y_{ji} t_{ji}$.

For the static capacity model the arrival rate $\lambda_j$ on basis job class is distributed over the machines and results in a utilization profile, that corresponds to the capacity usage. A proper distribution or resource allocation is therefore a center piece of capacity planning, it will be used to judge if the demand can be satisfied and to identify the bottlenecks. The capacity model selected for the investigations is based on the idea of minimizing the maximum utilization - that means, to minimize utilization on the bottleneck. This approach can be extended to a lexicographic mini-max objective like in Ogryczak [20] to identify all levels of equally utilized equipment (cf. resource pool concept in Gold [9, 10]). A solution that is optimal with respect to the lexicographic mini-max objective has the property that only machines of the same level may share the load of a common job class.

The LP defined in (1-4) is called the basic model, where the maximum load is minimized. According to (3) $\rho$ is an upper limit to the utilization $u_i$ for all tools $i \in I$ and the decision variable $x_{ji}$ corresponds to the time invested in processing lots from job class $j$ on machine $i$. The service rate $\mu_{ji} = \frac{1}{t_{ji}}$ corresponds to number of lots that can be processed in one time unit. Therefore the number of processed units of job class $j$ on machine $j$ is $y_{ji} = x_{ji}\mu_{ji}$. In order to satisfy the demand, constraint (2) has to be satisfied.

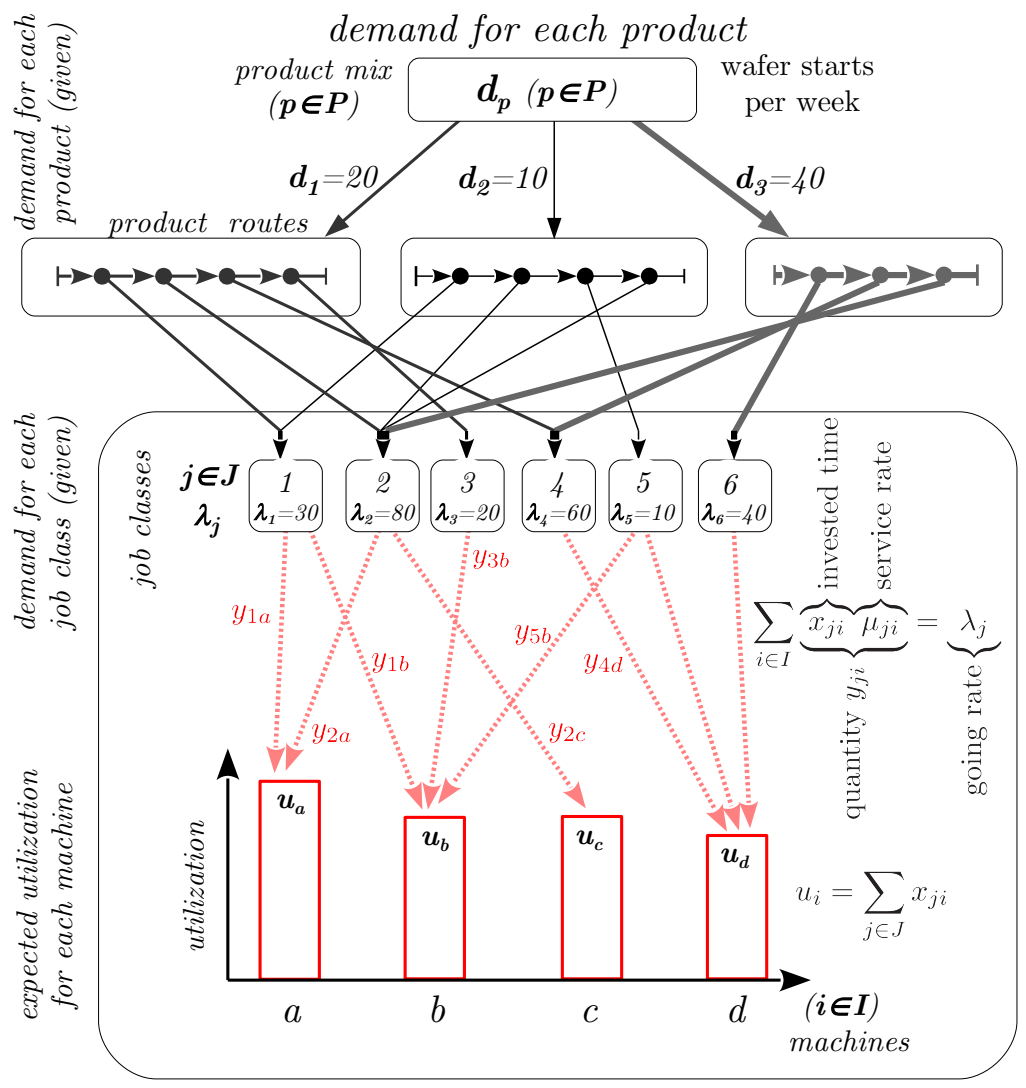

Figure 3: A simplified capacity planning process - the Figure is adapted from Romauch and Klemmt [23]. A given demand on the basis of products is decomposed into a demand for each job class. Finally a distribution of the workload results in a forecast for the expected utilization.

$$\min \rho \tag{1}$$

$$s.t. \quad \sum_{i\in I} x_{ji}\mu_{ji} = \lambda_j \quad (\forall j \in J), \tag{2}$$

$$u_i = \sum_{j\in J} x_{ji} \leq \rho \quad (\forall i \in I), \tag{3}$$

$$x_{ji} \geq 0. \tag{4}$$

## 3. Capacity Planning for Cluster Tools

The LP defined in (1-4) is a valid model for parallel single servers in heavy traffic (see e.g. Harrison and Lopez [12]), where each machine is represented by a queue with a given service rate.

The models presented in this paper are based on these assumptions and extend the model to Cluster-Tools. According to Lopez and Wood [18] the main operating modes are the serial and the parallel mode. A Cluster-Tool is operated in serial mode if each wafer visits all chambers in a sequence and it is working in parallel mode if each wafer visits only one chamber.

The serial mode can be approximated (lower bound) by separating the chambers of the Cluster-Tool from the mainframe. For instance, a Cluster-tool $i$ with three chambers $\{A, B, C\}$ is represented by the mainframe $i$, and the chambers $(i, A)$, $(i, B)$ and $(i, C)$. An activity $j$ calls chambers and mainframe at the same time. The corresponding service rates for the chambers are $\mu_{j,(i,A)}$, $\mu_{j,(i,B)}$ and $\mu_{j,(i,C)}$. Since the slowest process defines the throughput of the tool, hence the service rate for the mainframe is defined by $\mu_{j,i} = \min_{u\in A,B,C}\{\mu_{j,(i,v)}\}$. Therefore it is sufficient to add the following constraint to (1-4):

$$u_{i,v} = \sum_{j\in J} \frac{x_{ji}\mu_{ji}}{\mu_{j,(i,v)}} \leq \rho \quad (\forall i \in I, v \in \{A, B, C\}) \tag{5}$$

The objective of this paper is to propose a similar approximation for Cluster-Tools in parallel mode. The corresponding models need a bit more explanation. For Cluster-Tools in parallel mode the service rate depends on the selected subset of chambers $r\in \mathcal{R} =\{A, B, C, AB, ...\}$ that are selected to process a given type of lot. The subset of chambers is also called a recipe. The service time $\mu_{j,i,r}$ depends on the job class $j$, the Cluster tool $i$ and the recipe $r$. The demand is represented by the arrival rate $\lambda_j$.

The distribution of lots is determined by $x_{j,i,r}$, which is the time invested in activity $(j, i)$ with recipe $r$. In Ortner [21] a parallel Cluster-Tool model for two Load locks and three processing chambers is discussed. The basic idea is the identification of categories for optimal schedules (see Figure 4) where the makespan can be represented by a simple formula. Each formula is the weighted sums of the times invested in a given recipes amd can be written as $\sum_r w_{i,r}x_{i,r}$ where $x_{i,r} = \sum_j x_{j,i,r}$.

For instance, if all possible processes are running in parallel which means that the load locks are the limiting resource, then $x_{i,ABC}+\frac{x_{i,AB}+x_{i,AC}+x_{i,BC}+x_{i,A}+x_{i,B}+x_{i,C}}{2}$ is the corresponding make-span. If this is not the case, then a single chamber recipe exists that is too short or too long to run in parallel, which means that the make-span must be larger than that.

It was shown, that the make-span can be represented by the maximum value of five formulas. The corresponding proof in Ortner [21] is constructive and generates feasible schedules for each case identified.

The LP (6-9) is an abstraction of the LP found in Ortner [21], where the matrix $B$ is defined in Table 1.

In (6-9) we present an extension of (1-4) that can be used to represent the approach presented in Ortner [21]. For the three chamber case, the matrix $B$ represents the formulas given in Figure 4 and the corresponding values can be found in Table 1.

One of the contributions of the paper is to derive matrices for more than three chambers.

$$\min \rho \tag{6}$$

$$\sum_{\substack{i \in I, r \in \mathcal{R}: \\ (j,i,r) \in \mathcal{C}}} x_{j,i,r}\mu_{j,i,r} = \lambda_j \quad (\forall j \in J) \tag{7}$$

$$\sum_{j\in J, r\in\mathcal{R}} B_{k,r}x_{j,i,r} \leq \rho \quad (\forall i \in I, k \in K) \tag{8}$$

$$x_{j,i,r} \geq 0 \quad (j, i, r) \in \mathcal{C} \tag{9}$$

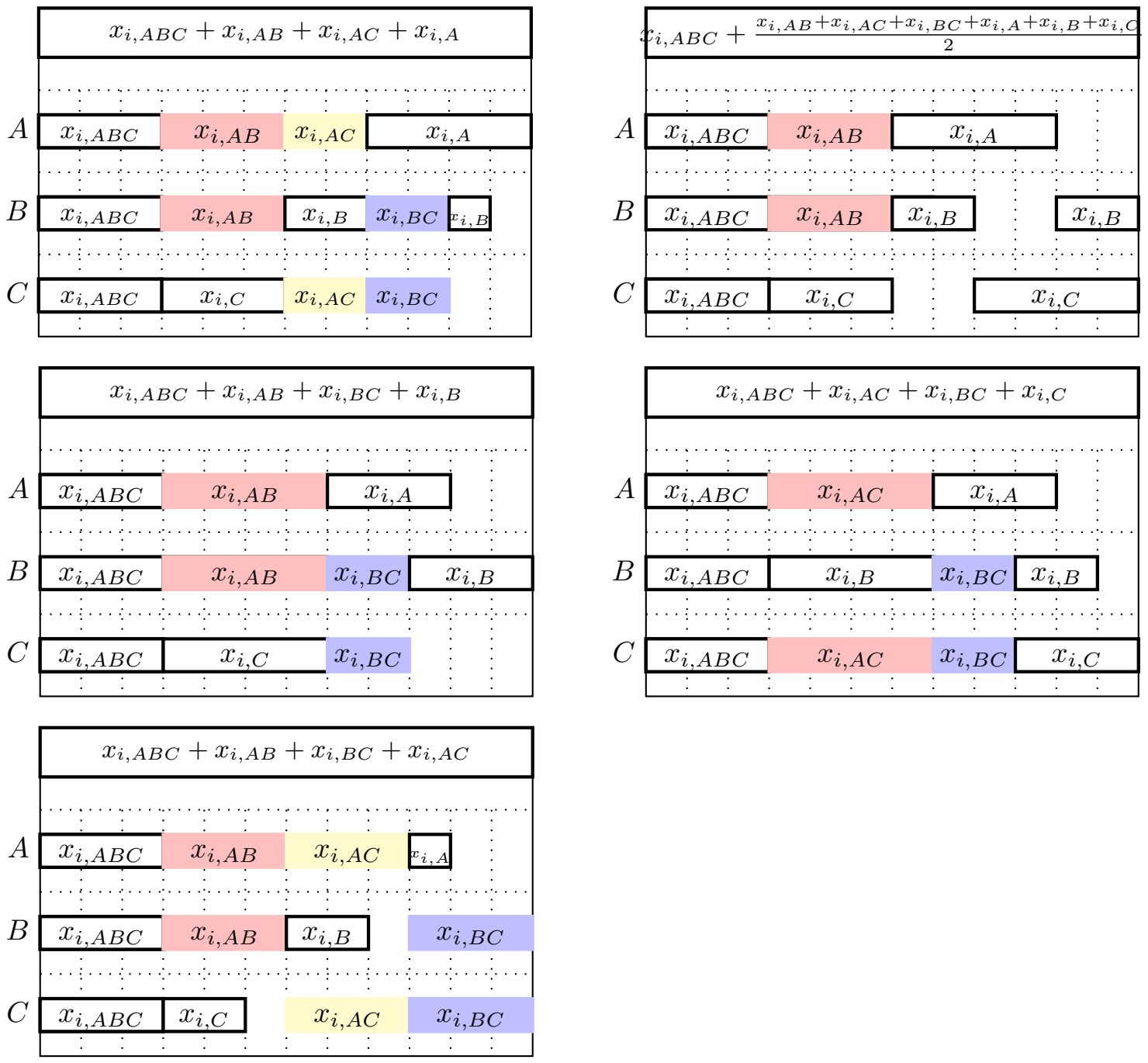

Figure 4: Illustration of different categories of optimal schedules. Five examples are given, one for each representative schedule. Each Figure ia a Gantt-chart (x-axis: time) on basis chambers. The formula in the header contains the formula to calculate the makespan, it represents this category. Two-chamber-recipes are colored to highlight the joint usage of two chambers.

Table 1: Matrix $B = (B_{k,r})$ for the dual based model (6-9) when considering three chambers.

| | r=$A$ | $B$ | $C$ | $AC$ | $BC$ | $AB$ | $ABC$ |
|---|---|---|---|---|---|---|---|
| k=1 | 0.5 | 0.5 | 0.5 | 0.5 | 0.5 | 0.5 | 1 |
| 2 | 1 | | | 1 | | 1 | 1 |
| 3 | | 1 | | | 1 | 1 | 1 |
| 4 | | | | 1 | 1 | 1 | 1 |
| 5 | | | 1 | 1 | 1 | | 1 |

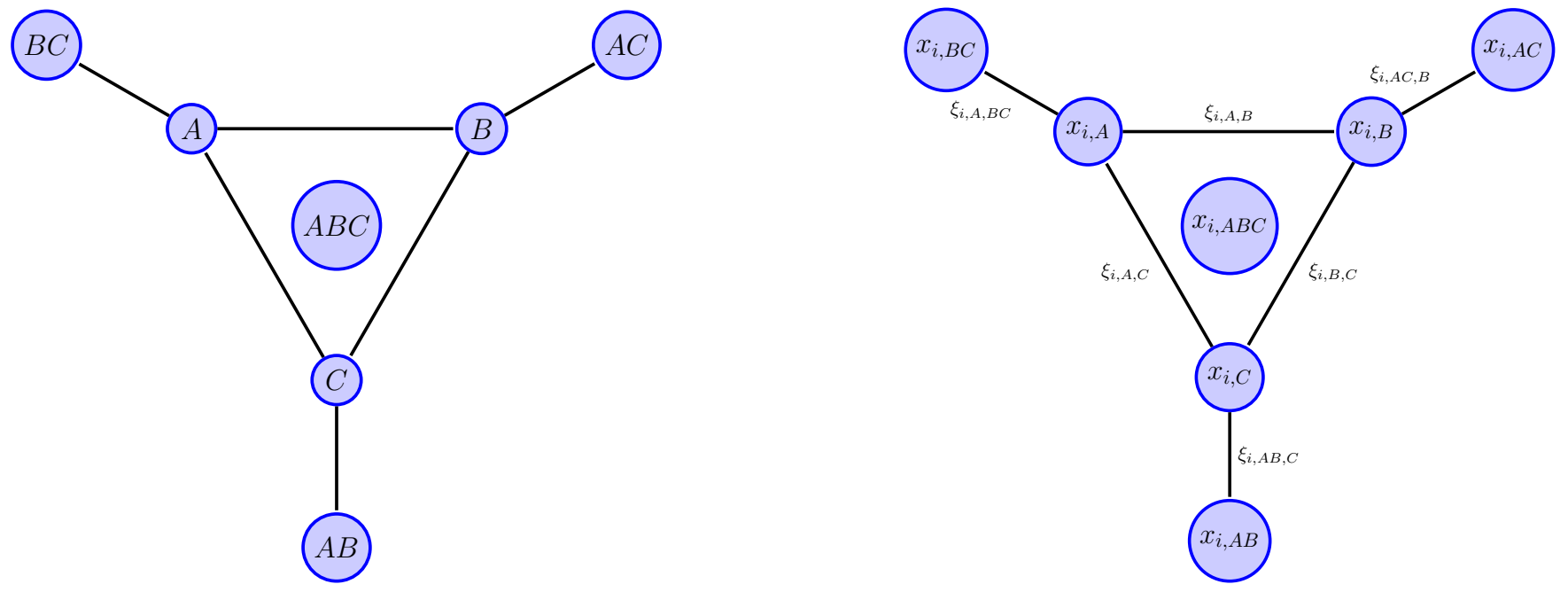

Figure 5: Parallel processing for three chambers, graph representation $G(\mathcal{R}, \mathcal{E})$. The vertices $\mathcal{R}$ represent the recipes and the edges $\mathcal{E}$ represent possible parallelizations. The Figure on the right hand side represents the parallelization opportunities for a given Cluster-tool $i$. For instance, the variable $\xi_{i,AB,C}$ defines the time invested in parallel processing of recipe $AB$ and recipe $C$.

Before tackling this problem, an alternative modeling approach that considers more than three chambers is presented.

### *3.1. An alternative modeling approach that exploits parallel processing*

For a single equipment $i$, where the time invested in certain recipes $x_{j,i,r}$ is given and no parallel processing is allowed (one load lock) the resulting makespan is $\sum_{j,r} x_{j,i,r}$. In case two load locks are available it is possible to exploit parallel processing. Since $x_{j,i,r}$ cannot be changed, the time spent in parallel processing is the only way to reduce the makespan. This parallelization problem can be seed as a sub problem and will be explained in this chapter. A similar idea was used in Tian et al. [26] for parallel processor scheduling.

In Figure 5 the graph representation $G(\mathcal{R}, \mathcal{E})$ of the parallelization problem for three chambers can be found. The vertices $\mathcal{R}$ represent the recipes and the edges $\mathcal{E}$ represent possible parallelizations. On the right hand side the parallelization opportunities for a given Cluster-tool $i$ are illustrated. For instance, when considering the parallelizations with recipe $C$, then there are three possible candidates $\{A, B, AB\}$. Each of these options is reflected by a variable that defines the time used for this recipe combination, i.e. $\xi_{i,A,C}$, $\xi_{i,B,C}$ and $\xi_{i,AB,C}$. Also note that the time invested in parallel processing is limited by the given values $x_{j,i,r}$. For instance when considering $\xi_{i,AB,C}$ two restrictions can be found: $\xi_{i,AB,C} \leq x_{i,AB}$ and $\xi_{i,AB,C} + \xi_{i,A,C} + \xi_{i,B,C} \leq x_{i,C}$. According to that, a parallelization that maximizes $\sum_{(r_1,r_2)\in\mathcal{E}} \xi_{i,r_1,r_2}$ solves the sub-problem.

The prallelization graph that considers four chambers is depicted in Figure 6 and the number of vertices is already twice as large. The graph is also not planar anymore, since it contains a $K_{3,3}$ given by $\{A, B, AB\} \times \{C, D, CD\}$.

The general formulation of the corresponding sub problem for optimal parallelization is represented by the following LP:

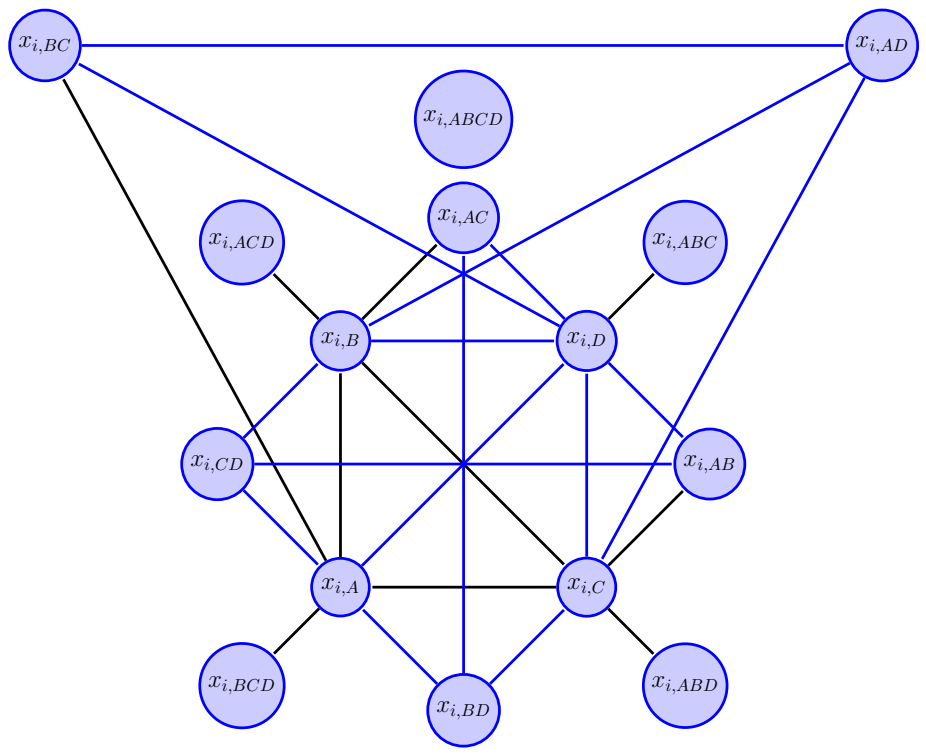

Figure 6: Parallel processing for four chambers, graph representation $G(\mathcal{R}, \mathcal{E})$.

$$\max \textstyle\sum_{(r_1,r_2)\in\mathcal{E}} \xi_{i,r_1,r_2} \tag{10}$$

$$\textstyle\sum_{r_2:(r,r_2)\in\mathcal{E}} \xi_{i,r,r_2} + \sum_{r_1:(r_1,r)\in\mathcal{E}} \xi_{i,r_1,r} \leq x_{i,r} \quad (\forall r \in \mathcal{R}) \tag{11}$$

$$\xi_{i,r_1,r_2} \geq 0 \quad (\forall (r_1,r_2) \in \mathcal{E}) \tag{12}$$

The sub-problem (10-12) be integrated in the master LP as follows:

$$\min \rho \tag{13}$$

$$\sum_{\substack{i \in I, r \in \mathcal{R}:\\ (j,i,r)\in\mathcal{C}}} x_{j,i,r}\mu_{j,i,r} = \lambda_j \quad j \in J \tag{14}$$

$$x_{i,r} = \textstyle\sum_j x_{j,i,r} \quad i \in I,\ r \in \mathcal{R} \tag{15}$$

$$\textstyle\sum_{r_2:(r,r_2)\in\mathcal{E}} \xi_{i,r,r_2} + \sum_{r_1:(r_1,r)\in\mathcal{E}} \xi_{i,r_1,r} \leq x_{i,r} \quad i \in I,\ r \in \mathcal{R} \tag{16}$$

$$\textstyle\sum_{r\in\mathcal{R}} x_{i,r} - \sum_{(r_1,r_2)\in\mathcal{E}} \xi_{i,r_1,r_2} \leq \rho \quad i \in I \tag{17}$$

$$x_{j,i,r} \geq 0 \quad (j,i,r) \in \mathcal{C} \tag{18}$$

$$x_{i,r} \geq 0 \quad i \in I,\ r \in \mathcal{R} \tag{19}$$

$$\xi_{i,r_1,r_2} \geq 0 \quad i \in I,\ (r_1,r_2) \in \mathcal{E} \tag{20}$$

In order to count the number of variables $x_{j,i,r}$, $x_{i,r}$ and $\xi_{i,r_1,r_2}$ it is sufficient to count the vertices and edges in the parallelization graph. Suppose that $n$ chambers are available, then $|\mathcal{R}| = 2^n - 1$ recipes are available. There are $\binom{n}{k}$ possibilities to choose a set of $k$ chambers and each partition defines non-conflicting recipe pair, that makes $\frac{2^k-2}{2}(= 2^{k-1} - 1)$ edges. Therefore, the number of edges is $|\mathcal{E}| = \sum_{k=0}^{n}(2^{k-1} - 1) = \frac{3^n-1}{2} - (2^n - 1)$ in total. According to that the coefficient matrix of (13-20) has $1+|\mathcal{C}| + |I|\frac{3^n-1}{2}$ columns, $|J| + |I|(1 + 2|\mathcal{R}|)$ rows and $2|\mathcal{C}| + |I|(1 + 3|\mathcal{R}| + 2|\mathcal{E}|)$ nonzeros.

*3.2. Analysis of the parallelization problem*

The second approach is based on investigating the dual of the sub-problem (10-12). The problem (10-12) can be represented by a max-flow problem which gives some insights about the structure of basic feasible solutions.

The corresponding max-flow problem is defined as follows, the set of vertices and edges and the capacity limits are defined by: $V = \mathcal{R} \cup \widetilde{\mathcal{R}} \cup \{s,t\}$, where $\widetilde{\mathcal{R}}$ is a copy of $\mathcal{R}$. The arcs between $\mathcal{R}$ and $\widetilde{\mathcal{R}}$ are a bipartite representation of $\mathcal{E}$, together with the links to the source, and the links to the sink, the set of arcs is defined by $\widetilde{\mathcal{E}}$, i.e $\widetilde{\mathcal{E}} = \{(r,\widetilde{r}) \in \mathcal{R} \times \widetilde{\mathcal{R}} : (r,\widetilde{r}) \in \mathcal{E}\} \cup \{(s,r) : r \in \mathcal{R}\} \cup \{(\widetilde{r},t) : \widetilde{r} \in \widetilde{\mathcal{R}}\}$. The arc capacities are infinite for all arcs except the links to source and the links to the sink. In Figure 7 the max-flow version for three chambers is depicted where the graph $G(\mathcal{R} \cup \widetilde{\mathcal{R}} \cup \{s,t\}, \widetilde{\mathcal{E}})$ is illustrated. The meaning of the flows $\eta_{i,r_1,\tilde{r_2}}$ and $\eta_{i,r_2,\tilde{r_1}}$ is directly linked to $\xi_{i,r_1,r_2}$ since the maximum flow formulation can be understood as a duplication of the original problem. The details will be elaborated in a transformation of the problem. Before that, the LP of the corresponding max-flow problem will be defined in (21-26). The objective (22) is to maximize the flow into the sink $t$. The constraints (22) and (23) represent the flow balance equations. The capacities on the arcs are formulated in (24) and (25).

$$\max \textstyle\sum_{(\widetilde{r},t)\in\widetilde{\mathcal{E}}} \eta_{i,\widetilde{r},t} \tag{21}$$

$$\textstyle\sum_{\widetilde{r}:\,(r,\widetilde{r})\in\widetilde{\mathcal{E}}} \eta_{i,r,\widetilde{r}} = \eta_{i,s,r} \qquad (\forall r \in \mathcal{R}) \tag{22}$$

$$\textstyle\sum_{r:\,(r,\widetilde{r})\in\widetilde{\mathcal{E}}} \eta_{i,r,\widetilde{r}} = \eta_{i,\widetilde{r},t} \qquad (\forall \widetilde{r} \in \widetilde{\mathcal{R}}) \tag{23}$$

$$\eta_{i,s,r} \leq \tfrac{x_{i,r}}{2} \qquad (\forall r \in \mathcal{R}) \tag{24}$$

$$\eta_{i,\widetilde{r},t} \leq \tfrac{x_{i,r}}{2} \qquad (\forall r \in \mathcal{R}) \tag{25}$$

$$\eta_{i,e} \geq 0 \qquad (\forall e \in \widetilde{\mathcal{E}}) \tag{26}$$

To see that the max-flow version is a suitable representation of the sub problem, a transformation of solutions will be established.

If $\eta$ is a feasible solution for (21-26), then $\xi$ defined by $\xi_{i,r_1,r_2} \leftarrow \eta_{i,r_1,\widetilde{r_2}} + \eta_{i,r_2,\widetilde{r_1}}$ is a feasible solution for (10-12) and the objective value is the same. In order to prove this statement it is necessary to prove that (11) is satisfied. The left hand side of (11) has the following form: $\sum_{r_2:(r,r_2)\in\mathcal{E}} (\underbrace{\eta_{i,r,\widetilde{r_2}}}_{(i.1)} + \overbrace{\eta_{i,r_2,\widetilde{r}}}^{(ii.1)}) + \sum_{r_1:(r_1,r)\in\mathcal{E}} (\overbrace{\eta_{i,r_1,\widetilde{r}}}^{(ii.2)} + \underbrace{\eta_{i,r,\widetilde{r_1}}}_{(i.2)})$. Note that $(r, r_2) \in \mathcal{E}$ is represented by two arcs $(r, \tilde{r_2}) \in \widetilde{\mathcal{E}}$ and $(r_2, \tilde{r}) \in \widetilde{\mathcal{E}}$, Therefore the components $(i.1)$ and $(i.2)$ can be combined and because of (22,24) the corresponding part is bounded by $\sum_{r_2:(r,r_2)\in\mathcal{E}} \eta_{i,r,\widetilde{r_2}} + \sum_{r_1:(r_1,r)\in\mathcal{E}} \eta_{i,r,\widetilde{r_1}} = \sum_{\widetilde{r}:(r,\widetilde{r})\in\widetilde{\mathcal{E}}} \eta_{i,r,\widetilde{r}} = \eta_{i,s,r} \leq \frac{x_{i,r}}{2}$. Because of (23,25) the part that corresponds to $(ii.1)$ and $(ii.2)$ is bounded by $\sum_{r_2:(r,r_2)\in\mathcal{E}} \eta_{i,r_2,\widetilde{r}} + \sum_{r_1:(r_1,r)\in\mathcal{E}} \eta_{i,r_1,\widetilde{r}} = \sum_{r:(r,\widetilde{r})\in\widetilde{\mathcal{E}}} \eta_{i,r,\widetilde{r}} = \eta_{i,\widetilde{r},t} \leq \frac{x_{i,r}}{2}$. Hence, $\xi$ is feasible for (10-12).

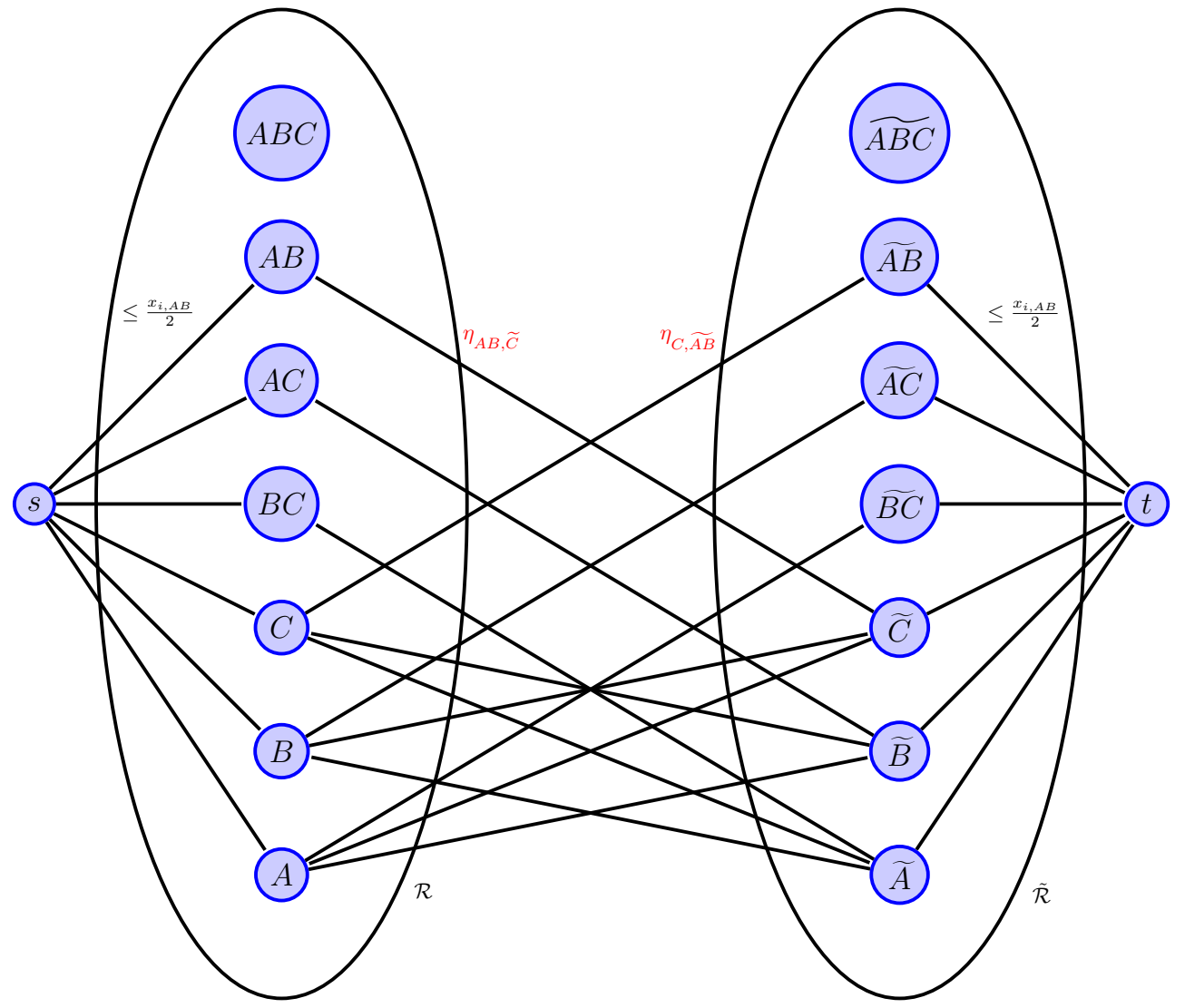

Figure 7: Graph representation $G(\mathcal{R} \cup \widetilde{\mathcal{R}} \cup \{s,t\}, \widetilde{\mathcal{E}})$ of the maximum flow formulation of the sub problem for three chambers. The flow can be interpreted as the time used for parallel processing: $\xi_{AB,C} \leftarrow \eta_{C,\widetilde{AB}} + \eta_{AB,\widetilde{C}}$.

The dual is therefore a min-cut problem and for the solution space it is sufficient to consider selecting arcs that are connected to the sink or to the source - these cuts are called *basic cuts*. In other words, for every cut it is possible to construct a basic cut that has a cut value that is smaller or equal. The formulation of the dual of the maximum flow formulation (21-26) can be found in (27-30). The dual variables are called $\pi_{i,r}$ and $\widetilde{\pi}_{i,r}$. The constraint (28) reveals the vertex covering aspect of the problem.

$$\min \sum_{r\in\mathcal{R}} x_{i,r} \frac{\pi_{i,r}+\widetilde{\pi}_{i,r}}{2} \tag{27}$$

$$\pi_{i,r_1} + \widetilde{\pi}_{i,r_2} \geq 1 \qquad \forall (r_1, \widetilde{r_2}) \in \widetilde{\mathcal{E}} \tag{28}$$

$$\pi_{i,r} \geq 0 \qquad \forall r \in \mathcal{R} \tag{29}$$

$$\widetilde{\pi}_{i,r} \geq 0 \qquad \forall r \in \mathcal{R} \tag{30}$$

This problem is totally uni-modular, because the columns can be partitioned according to the criterion proposed in Heller and Tompkins [13].

For the search space, it is sufficient to consider basic cuts. A basic cut $\mathcal{B}$ can be represented by the corresponding nodes that are selected, $\mathcal{B} \subset \mathcal{R} \cup \widetilde{\mathcal{R}}$ and each arc $(r_1, \widetilde{r_2})$ between $\mathcal{R}$ and $\widetilde{\mathcal{R}}$ is covered by $\mathcal{B}$ to satisfy (28). Furthermore, it is sufficient to consider basic cut that cannot be reduced by eliminating a node, that means that by definition, removing one node of a *minimal basic cut* leads to an infeasible solution.

Therefore the complement ($\mathcal{B}^c = \mathcal{R} \cup \widetilde{\mathcal{R}} \setminus \mathcal{B}$) of a minimal basic cut is a maximum independent set. Efficient algorithm for listing all maximum independent sets are known (see Kashiwabara et al. [17]) and can be used to find all *minimal basic cuts.*

Suppose that a listing $\{(\pi^k_{i,r}, \widetilde{\pi}^k_{i,r})\}_{k\in\mathcal{K}}$ that represents all minimal basic cuts is given, then the optimal value for (27-30) corresponds to the cut value with the smallest value in the listing of all *minimal basic cuts.*

Suppose that all minimal basic cuts $\{(\pi^k_{i,r}, \widetilde{\pi}^k_{i,r})\}_{k\in\mathcal{K}}$ are given, then the smallest value $\sum_{r\in\mathcal{R}} x_{i,r}\left(\frac{\pi^k_{i,r}+\widetilde{\pi}^k_{i,r}}{2}\right)$ defines the optimum.

In order to remove symmetric solutions, the solutions will be identified by vectors $(\Pi_{i,r,k})$ where $\left(\Pi_{i,r,k} \leftarrow \frac{\pi^k_{i,r}+\widetilde{\pi}^k_{i,r}}{2}\right)$, hence the optimum is defined by $\min_k \sum_{r\in\mathcal{R}} x_{i,r}\Pi_{i,r,k}$. The minimum of a set of numbers is defined as the largest lower bound of these numbers, therefore the optimum can be represented by the LP formulated in (31-32), where constraint (32) ensures that $\chi$ is a lower bound of the cut values and the objective (31) defines that $\chi$ is the largest lower bound:

$$\max \chi \tag{31}$$

$$\chi \leq \sum_{r\in\mathcal{R}} x_{i,r}\Pi_{i,r,k} \quad \forall k \in \mathcal{K} \tag{32}$$

Therefore the shortest possible makespan for given $x_{i,r}$ can be obtained by changing the the objective (32) to $\min \sum_{r\in\mathcal{R}} x_{i,r} - \chi$. A substitution of $\chi$ with $\rho \leftarrow \sum_{r\in\mathcal{R}} x_{i,r} - \chi$ in (33-32) leads to the following equivalent LP:

$$\min \quad \rho \tag{33}$$

$$\sum_{r\in\mathcal{R}} (1 - \Pi_{i,r,k})\, x_{i,r} \leq \rho \quad \forall k \in \mathcal{K} \tag{34}$$

The new formulation of the sub problem will be integrated in the master problem (13-20) by replacing the constraints (17,16) with (34). The resulting master LP is formulated in (3540):

$$\min \rho \tag{35}$$

$$\sum_{\substack{i\in I, r\in\mathcal{R}:\\ (j,i,r)\in\mathcal{C}}} x_{j,i,r}\mu_{j,i,r} = \lambda_j \quad j \in J \tag{36}$$

$$x_{i,r} = \sum_j x_{j,i,r} \quad i \in I,\, r \in \mathcal{R} \tag{37}$$

$$\sum_{r\in\mathcal{R}}(1 - \Pi_{i,r,k})x_{i,r} \leq \rho \quad i \in I, k \in \mathcal{K} \tag{38}$$

$$x_{j,i} \geq 0 \quad i \in I,\, r \in \mathcal{R} \tag{39}$$

$$x_{j,i,r} \geq 0 \quad (j,i,r) \in \mathcal{C} \tag{40}$$

The problem (27-30) can be interpreted as a relaxation of the vertex cover problem, therefore the basic solutions are half integral (compare Hochbaum [14], type 2var).

If the number of chambers is fixed, the coefficients $(1 - \Pi_{i,r,k})$ only depend on the number of chambers, therefore a representation by a $\mathcal{R} \times \mathcal{K}$ matrix $1-\Pi = (1-\Pi_{r,k})$ can be used. In order to reduce the size of the matrix, all rows of $1 - \Pi$ that are redundant will be removed. The details to the corresponding Algorithm can be found in AppendixA.

For three chambers the model (3540) coincides with the model presented in Ortner [21] (compare (6-9)) and the matrix $(1 - \Pi)$ is equivalent to the matrix presented in Table 1. Therefore, the presented methodology replaces (6-9) and provides an alternative to the elementary proof given in Ortner [21].

More importantly, the approach reveals a generalization of the model in Ortner [21] to more than three chambers. For instance, the matrix for four chambers is given in Table 2. Note that line 1 corresponds to complete parallelization, while line 8 corresponds to the sum of all recipes that contain $A$ and line 12 corresponds to chamber $B$, line 18 represents the utilization for chamber $C$ and line 23 represents the utilization for chamber $D$. Note that the coefficient matrix of (3540) has $1+|\mathcal{C}|+|I|\frac{3^n-1}{2}$ columns, $|J|+|I|(|\mathcal{R}|+|\mathcal{K}|)$ rows and $2|\mathcal{C}| + |I|(|\mathcal{R}| + 1 + |\mathcal{K}| + |1 - \Pi|)$ nonzeros, where $|1 - \Pi|$ denotes the number of nonzeros in $(1 - \Pi_{i,r,k})$. Known values for $|\mathcal{K}|$ and $|1 - \Pi|$ can be found in Table 4. According to that, the generalized model for five chamber is already getting fairly large.

In the next section a computational experiment will be used to compare the new approach and the generalized model.

## 4. Computational Experiments

In this Section the computational results for the generalized formulation (35-40) and the alternative model (13-20) for the a large set of benchmark instances (http://www.univie.ac.at/prolog/research/Clustertool/) is presented. The instances were solved with Gurobi 6.0 on a desktop computer with a i7-2670QM@2.2 GHz processor.

The structure of the generalized formulation (13-20) seems to be much simpler than the alternative model (35-40), but especially for more than three chambers the coefficient matrix of the generalized formulation tends to grow faster and therefore it is interesting to examine which model performs better. The size of the problems is indicated in Table 5 and Table 7, where the number of nonzero entries in the coefficient matrix of the alternative model and the scaling factor for the generalized model are documented (all chamber released).

The number of nonzeros of the coefficient matrix for the generalized model is $nonzero_{gen} = 2|\mathcal{C}|+|I|(|\mathcal{R}|+1+|\mathcal{K}|+|\Pi|)$ and the for the alternative formulation it is $nonzero_{alt} = 2|\mathcal{C}| + |I|(1 + 3|\mathcal{R}| + 2|\mathcal{E}|)$ . Therefore the difference of the $\delta = nonzero_{gen} - nonzero_{alt}$ can be calculated as follows:

$\delta = |I|(|\mathcal{K}| + |\Pi| - 2(|\mathcal{R}| + |\mathcal{E}|)) = |I|(1 + |\mathcal{K}| + |\Pi| - 3^n)$

Table 2: Matrix $B$ for the dual based model (6-9) when considering four chambers.

| | $A$ | AB | ABC | ABCD | ABD | AC | ACD | AD | B | BC | BCD | BD | C | CD | D |
|---|---|---|---|---|---|---|---|---|---|---|---|---|---|---|---|
| 1 | 0.5 | 0.5 | 0.5 | 1 | 0.5 | 0.5 | 0.5 | 0.5 | 0.5 | 0.5 | 0.5 | 0.5 | 0.5 | 0.5 | 0.5 |
| 2 | 0.5 | 0.5 | 1 | 1 | 0.5 | 0.5 | 0.5 | 0.5 | 0.5 | 0.5 | 0.5 | 0.5 | 0.5 | 0.5 | |
| 3 | 0.5 | 0.5 | 0.5 | 1 | 1 | 0.5 | 0.5 | 0.5 | 0.5 | 0.5 | 0.5 | 0.5 | | 0.5 | 0.5 |
| 4 | 0.5 | 0.5 | 1 | 1 | 1 | 0.5 | 0.5 | 0.5 | 0.5 | 0.5 | 0.5 | 0.5 | | 0.5 | |
| 5 | 0.5 | 0.5 | 0.5 | 1 | 0.5 | 0.5 | 1 | 0.5 | | 0.5 | 0.5 | 0.5 | 0.5 | 0.5 | 0.5 |
| 6 | 0.5 | 0.5 | 1 | 1 | 0.5 | 0.5 | 1 | 0.5 | | 0.5 | 0.5 | 0.5 | 0.5 | 0.5 | |
| 7 | 0.5 | 0.5 | 0.5 | 1 | 1 | 0.5 | 1 | 0.5 | | 0.5 | 0.5 | 0.5 | | 0.5 | 0.5 |
| 8 | 1 | 1 | 1 | 1 | 1 | 1 | 1 | 1 | | | | | | | |
| 9 | | 0.5 | 0.5 | 1 | 0.5 | 0.5 | 0.5 | 0.5 | 0.5 | 0.5 | 1 | 0.5 | 0.5 | 0.5 | 0.5 |
| 10 | | 0.5 | 1 | 1 | 0.5 | 0.5 | 0.5 | 0.5 | 0.5 | 0.5 | 1 | 0.5 | 0.5 | 0.5 | |
| 11 | | 0.5 | 0.5 | 1 | 1 | 0.5 | 0.5 | 0.5 | 0.5 | 0.5 | 1 | 0.5 | | 0.5 | 0.5 |
| 12 | | 1 | 1 | 1 | 1 | | | | 1 | 1 | 1 | 1 | | | |
| 13 | | 0.5 | 0.5 | 1 | 0.5 | 0.5 | 1 | 0.5 | | 0.5 | 1 | 0.5 | 0.5 | 0.5 | 0.5 |
| 14 | | 1 | 1 | 1 | 1 | 1 | 1 | | | 1 | 1 | | | | |
| 15 | | 1 | 1 | 1 | 1 | 1 | 1 | 1 | | | 1 | | | | |
| 16 | | 1 | 1 | 1 | 1 | | 1 | | | 1 | 1 | 1 | | | |
| 17 | | 1 | 1 | 1 | 1 | | 1 | 1 | | | 1 | 1 | | | |
| 18 | | | 1 | 1 | | 1 | 1 | | | 1 | 1 | | 1 | 1 | |
| 19 | | | 1 | 1 | 1 | 1 | 1 | | | 1 | 1 | | | 1 | |
| 20 | | | 1 | 1 | 1 | 1 | 1 | 1 | | | 1 | | | 1 | |
| 21 | | | 1 | 1 | 1 | | 1 | | | 1 | 1 | 1 | | 1 | |
| 22 | | | 1 | 1 | 1 | | 1 | 1 | | | 1 | 1 | | 1 | |
| 23 | | | | 1 | 1 | | 1 | 1 | | | 1 | 1 | | 1 | 1 |

Table 4: Known values for $|\mathcal{K}|$ and $|1-\Pi|$

| | n=1 | 2 | 3 | 4 | 5 |
|---|---|---|---|---|---|
| $\lvert\mathcal{K}\rvert$ | 1 | 2 | 5 | 23 | 590 |
| $\lvert 1-\Pi\rvert$ | 1 | 4 | 22 | 245 | 13740 |

Table 5: Number of nonzero entries in the coefficient matrix of the alternative model; unit: thousands.

| | | three chambers | | | | four chambers | | | | five chambers | | | |
|---|---|---|---|---|---|---|---|---|---|---|---|---|---|
| density | shape | sizecat=0 | 1 | 2 | 3 | 0 | 1 | 2 | 3 | 0 | 1 | 2 | 3 |
| 1 | 1:1 | 3.4 | 11.3 | 40.9 | 156.8 | 8.0 | 25.7 | 90.6 | 341.7 | 18.9 | 57.7 | 196.3 | 724.4 |
| | 1:4 | 2.2 | 8.7 | 32.9 | 123.7 | 5.1 | 19.4 | 71.8 | 267.9 | 11.6 | 42.4 | 153.0 | 562.8 |
| | 16:1 | 6.2 | 17.2 | 54.1 | 188.2 | 16.1 | 42.5 | 127.3 | 425.8 | 42.5 | 106.0 | 299.5 | 953.0 |
| | 4:1 | 4.4 | 13.8 | 46.7 | 171.7 | 10.8 | 32.3 | 105.8 | 379.2 | 26.8 | 75.9 | 236.9 | 820.2 |
| 2 | 1:1 | 4.3 | 15.3 | 59.2 | 232.5 | 9.9 | 34.2 | 129.8 | 504.0 | 22.7 | 75.3 | 277.3 | 1059.7 |
| | 1:4 | 3.0 | 10.7 | 42.2 | 165.5 | 6.9 | 23.6 | 91.9 | 357.5 | 15.3 | 51.1 | 194.5 | 747.9 |
| | 16:1 | 7.0 | 21.0 | 74.0 | 271.4 | 17.9 | 50.6 | 169.9 | 604.1 | 46.0 | 122.8 | 387.6 | 1321.4 |
| | 4:1 | 5.3 | 18.4 | 66.6 | 256.1 | 12.7 | 42.2 | 148.3 | 560.0 | 30.7 | 96.3 | 324.8 | 1193.8 |
| 3 | 1:1 | 4.5 | 17.4 | 68.7 | 271.7 | 10.4 | 38.7 | 149.9 | 587.8 | 23.8 | 84.6 | 319.0 | 1233.0 |
| | 1:4 | 3.2 | 11.5 | 46.0 | 181.4 | 7.2 | 25.4 | 99.9 | 391.4 | 16.1 | 54.8 | 211.0 | 818.1 |
| | 16:1 | 7.1 | 23.9 | 85.4 | 318.7 | 18.0 | 56.8 | 194.4 | 705.4 | 46.4 | 135.6 | 438.2 | 1530.8 |
| | 4:1 | 5.8 | 20.3 | 78.4 | 302.7 | 13.8 | 46.3 | 173.6 | 659.9 | 33.0 | 104.8 | 376.9 | 1400.2 |

Table 7: Scaling factor for nonzero entries in the coefficient matrix of the generalized model relative to alternative model. An entry of 2 means that the number of nonzero elements is twice as large.

| | | three chambers | | | | four chambers | | | | five chambers | | | |
|---|---|---|---|---|---|---|---|---|---|---|---|---|---|
| density | shape | sizecat=0 | 1 | 2 | 3 | 0 | 1 | 2 | 3 | 0 | 1 | 2 | 3 |
| 1 | 1:1 | 1.01 | 1.00 | 1.00 | 1.00 | 1.47 | 1.29 | 1.17 | 1.09 | 15.93 | 10.76 | 6.74 | 4.11 |
| | 1:4 | 1.00 | 1.00 | 1.00 | 1.00 | 1.38 | 1.22 | 1.12 | 1.06 | 13.44 | 8.49 | 5.06 | 3.13 |
| | 16:1 | 1.00 | 1.00 | 1.00 | 1.00 | 1.36 | 1.19 | 1.10 | 1.05 | 12.85 | 7.66 | 4.53 | 2.83 |
| | 4:1 | 1.00 | 1.00 | 1.00 | 1.00 | 1.37 | 1.19 | 1.10 | 1.06 | 13.17 | 7.64 | 4.68 | 3.00 |
| 2 | 1:1 | 1.00 | 1.00 | 1.00 | 1.00 | 1.27 | 1.16 | 1.08 | 1.04 | 10.21 | 6.52 | 3.90 | 2.51 |
| | 1:4 | 1.00 | 1.00 | 1.00 | 1.00 | 1.26 | 1.15 | 1.08 | 1.04 | 9.75 | 6.14 | 3.67 | 2.38 |
| | 16:1 | 1.01 | 1.01 | 1.01 | 1.00 | 1.93 | 1.71 | 1.47 | 1.28 | 27.55 | 22.26 | 16.05 | 10.46 |
| | 4:1 | 1.01 | 1.01 | 1.00 | 1.00 | 1.84 | 1.59 | 1.35 | 1.20 | 25.48 | 19.35 | 12.63 | 7.82 |
| 3 | 1:1 | 1.01 | 1.01 | 1.00 | 1.00 | 1.83 | 1.53 | 1.31 | 1.17 | 25.31 | 17.63 | 11.29 | 6.89 |
| | 1:4 | 1.01 | 1.01 | 1.00 | 1.00 | 1.70 | 1.47 | 1.28 | 1.16 | 22.00 | 15.86 | 10.51 | 6.50 |
| | 16:1 | 1.01 | 1.00 | 1.00 | 1.00 | 1.59 | 1.36 | 1.20 | 1.11 | 19.33 | 12.70 | 7.94 | 4.78 |
| | 4:1 | 1.01 | 1.00 | 1.00 | 1.00 | 1.55 | 1.32 | 1.17 | 1.09 | 18.09 | 11.76 | 6.98 | 4.22 |

Table 9: Auxiliary parameters $\delta_n$ and $\gamma_n$ for calculating $\delta = nonzeros_{gen} - nonzeros_{alt} = |I|\delta_n$ and $nonzero_{alt} = 2|\mathcal{C}| + |I|\gamma_n$ .

| | n=1 | 2 | 3 | 4 | 5 |
|---|---|---|---|---|---|
| $\delta_n$ | 0 | -2 | 1 | 188 | 14088 |
| $\gamma_n = \frac{3^{n+1}-2^{n+1}+1}{2}$ | 3 | 10 | 33 | 106 | 333 |

In Table 9 some values for $\frac{\delta}{|I|}$ can be found.

The instances that were used in the computational experiment vary in size and structure and the following four factors were controlling the construction:

- **sizecat** $\{0, 1, 2, 3\}$
  where, $|I| \cdot |J| = 100 \cdot 4^{sizecat}$ = 400,1600, 6400, 25600

- **shape** $\{1 : 4, 1 : 1, 4 : 1, 8 : 1\}$
  where $|I| : |J| = shape$, the proportion of tools and jobs

- **locked** $\{0, 3, 6, 9\}$
  *locked chambers:* $\frac{locked}{10}$ is the probability that a chamber is not available.

- **density** {1,2,3}
  1=sparse, 2=medium, 3=dense

For the three chamber model, the corresponding instances have up to 300.000 nonzeros and for four chamber up to 700.000 nonzeros for the case where all chambers are available. According to Table 10 the alternative model is significantly faster for four chambers, but in average slower for three chambers.

Table 10: Speedup factor for alternative model versus generalized model (detailed results). The bold entries represent cases were the generalized model is faster.

| | | | three chambers | | | | | four chambers | | | | |
|---|---|---|---|---|---|---|---|---|---|---|---|---|
| sizecat | shape | density | locked=0 | 3 | 6 | 9 | avg | locked=0 | 3 | 6 | 9 | avg |
| 0 | 1:4 | 1 | **0.84** | **0.82** | **0.88** | **0.8** | **0.84** | 1.25 | 1.12 | **0.83** | 1.11 | 1.08 |
| | | 2 | 1.07 | **0.81** | **0.61** | **0.93** | **0.86** | 1.2 | 1.15 | 1.33 | 1.18 | 1.22 |
| | | 3 | **0.83** | **0.85** | **0.91** | **0.93** | **0.88** | 1.2 | 1.21 | 1.27 | 1.22 | 1.23 |
| | 1:1 | 1 | **0.97** | **0.83** | **0.8** | **0.86** | **0.87** | **0.92** | **0.82** | **0.71** | 1.22 | **0.92** |
| | | 2 | **0.79** | **0.97** | **0.88** | **0.9** | **0.89** | 1.11 | 1.12 | 1.23 | 1.14 | 1.15 |
| | | 3 | **0.86** | **0.9** | **0.92** | **0.82** | **0.88** | **0.93** | 1.1 | 1.09 | 1.16 | 1.07 |
| | 4:1 | 1 | **0.67** | **0.65** | **0.81** | **0.68** | **0.7** | **0.66** | **0.79** | **0.82** | **0.8** | **0.77** |
| | | 2 | **0.96** | **0.73** | **0.77** | **0.76** | **0.81** | **0.77** | **0.82** | **0.82** | **0.93** | **0.84** |
| | | 3 | **0.83** | **0.74** | **0.77** | **0.71** | **0.76** | **0.82** | **0.8** | **0.83** | **0.9** | **0.84** |
| | 16:1 | 1 | **0.72** | **0.6** | **0.63** | **0.51** | **0.62** | **0.74** | **0.6** | **0.58** | **0.59** | **0.63** |
| | | 2 | **0.63** | **0.58** | **0.65** | **0.68** | **0.64** | **0.78** | **0.72** | **0.6** | **0.67** | **0.69** |
| | | 3 | **0.6** | **0.58** | **0.59** | **0.52** | **0.57** | **0.81** | **0.71** | **0.66** | **0.69** | **0.72** |
| 1 | 1:4 | 1 | **0.98** | **0.84** | **0.83** | **0.78** | **0.86** | 1.63 | 1.33 | 1.37 | 1.48 | 1.45 |
| | | 2 | **0.89** | **0.91** | **0.96** | **0.96** | **0.93** | 2.29 | 1.73 | 1.62 | 1.61 | 1.81 |
| | | 3 | **0.76** | **0.75** | **0.94** | **0.97** | **0.86** | 1.89 | 1.78 | 1.52 | 1.6 | 1.7 |
| | 1:1 | 1 | 1.08 | 0.92 | 1.02 | **0.85** | **0.97** | 1.43 | 1.43 | 1.42 | 1.34 | 1.41 |
| | | 2 | 1.11 | 0.82 | 1.09 | 1.01 | 1.01 | 1.53 | 1.48 | 1.44 | 1.43 | 1.47 |
| | | 3 | 1.11 | 1.07 | 1.1 | 1.73 | 1.25 | 1.83 | 1.43 | 1.46 | 1.42 | 1.54 |
| | 4:1 | 1 | **0.8** | **0.7** | **0.84** | **0.62** | **0.74** | 1.22 | 1.07 | **0.98** | 1.11 | 1.1 |
| | | 2 | 1.21 | **0.54** | 1.15 | **0.86** | **0.94** | 1.52 | 1.41 | 1.21 | 1.22 | 1.34 |
| | | 3 | 1.18 | **0.8** | **0.86** | **0.79** | **0.91** | 1.49 | 1.49 | 1.24 | 1.38 | 1.4 |
| | 16:1 | 1 | **0.66** | **0.57** | **0.71** | **0.68** | **0.66** | **0.99** | 1.04 | **0.83** | **0.84** | **0.96** |
| | | 2 | **0.68** | **0.6** | **0.48** | **0.72** | **0.62** | 1.29 | 1.27 | 1.12 | **0.94** | 1.16 |
| | | 3 | **0.74** | **0.51** | **0.7** | **0.76** | **0.68** | 1.26 | 1.27 | 1.14 | 1.07 | 1.19 |
| 2 | 1:4 | 1 | **0.91** | **0.97** | **0.74** | 1.02 | **0.91** | 2.05 | 2.32 | 1.92 | 1.82 | 2.03 |
| | | 2 | 1.08 | 1.06 | **0.98** | **0.96** | 1.02 | 2.22 | 2.02 | 2.03 | 2.05 | 2.08 |
| | | 3 | 1.06 | 1.09 | 1.06 | 1.09 | 1.08 | 2.04 | 2.02 | 1.87 | 1.98 | 1.98 |
| | 1:1 | 1 | 1.3 | 1.13 | **0.95** | 1.29 | 1.17 | 2.35 | 1.68 | 1.81 | 1.62 | 1.87 |
| | | 2 | 1.19 | 1.09 | 1.22 | 1.05 | 1.14 | 2.08 | 2.35 | 1.94 | 1.8 | 2.04 |
| | | 3 | 1.26 | 1.4 | 1.11 | 1.05 | 1.21 | 1.87 | 1.9 | 1.84 | 1.94 | 1.89 |
| | 4:1 | 1 | **0.95** | **0.89** | **0.88** | **0.92** | **0.91** | 1.69 | 1.78 | 1.66 | 1.53 | 1.67 |
| | | 2 | 1 | **0.91** | **0.89** | **0.93** | **0.93** | 1.68 | 1.7 | 1.83 | 1.7 | 1.73 |
| | | 3 | 2.43 | **0.99** | **0.94** | 1 | 1.34 | 1.82 | 1.76 | 1.78 | 1.69 | 1.76 |
| | 16:1 | 1 | **0.74** | **0.78** | **0.75** | **0.98** | **0.81** | 1.44 | 1.72 | 1.41 | 1.27 | 1.46 |
| | | 2 | **0.84** | **0.82** | **0.83** | **0.85** | **0.84** | 1.5 | 1.71 | 1.63 | 1.51 | 1.59 |
| | | 3 | **0.85** | **0.8** | **0.89** | **0.96** | **0.88** | 1.48 | 1.67 | 1.68 | 1.53 | 1.59 |
| 3 | 1:4 | 1 | 1.41 | 1.02 | 1.19 | 1.19 | 1.34 | 2.5 | 2.44 | 2.17 | 2 | 2.28 |
| | | 2 | 1.24 | 1.16 | 1.1 | **0.94** | 1.11 | 2.41 | 2.42 | 2.74 | 2.23 | 2.45 |
| | | 3 | 1.22 | 1.1 | **0.99** | 1.2 | 1.13 | 2.33 | 2.2 | 2.93 | 2.1 | 2.39 |
| | 1:1 | 1 | 1.77 | 1.32 | 1.05 | 1.07 | 1.3 | 1.82 | 2.26 | 2.04 | 1.94 | 2.02 |
| | | 2 | 1.28 | 1.48 | 1.33 | 1.05 | 1.29 | 2.94 | 2.24 | 2.13 | 2.11 | 2.36 |
| | | 3 | 1.33 | 1.31 | 1.32 | 1.09 | 1.26 | 3.09 | 3.52 | 2.21 | 2.11 | 2.73 |
| | 4:1 | 1 | 1.08 | **0.92** | **0.95** | 1.01 | **0.99** | 1.7 | 1.81 | 1.87 | 1.82 | 1.8 |
| | | 2 | 1.04 | **0.92** | 1 | 1.03 | 1 | 1.85 | 2 | 1.91 | 1.87 | 1.91 |
| | | 3 | 2.7 | 1.11 | 1 | 1.1 | 1.48 | 2.03 | 1.91 | 1.82 | 1.96 | 1.93 |
| | 16:1 | 1 | 1.03 | **0.83** | **0.89** | **0.92** | **0.92** | 1.63 | 1.91 | 1.88 | 1.62 | 1.76 |
| | | 2 | 1.02 | **0.94** | **0.92** | 1.01 | **0.97** | 1.7 | 1.7 | 1.94 | 1.74 | 1.77 |
| | | 3 | 1.16 | **0.91** | **0.95** | **0.99** | 1 | 1.63 | 1.64 | 1.88 | 1.79 | 1.74 |
| avg | | | 1.11 | **0.9** | **0.91** | **0.93** | **0.97** | 1.55 | 1.59 | 1.52 | 1.47 | 1.54 |

Table 12: CPU time in milliseconds.

| | | | three chambers | | | | | four chambers | | | | |
|---|---|---|---|---|---|---|---|---|---|---|---|---|
| sizecat | shape | density | locked=0 | 3 | 6 | 9 | *avg* | locked=0 | 3 | 6 | 9 | *avg* |
| 0 | 1:4 | 1 | 25 | 28 | 25 | 25 | 25.75 | 48 | 41 | 48 | 35 | 43.00 |
| | | 2 | 30 | 36 | 41 | 27 | 33.50 | 60 | 60 | 55 | 55 | 57.50 |
| | | 3 | 29 | 33 | 32 | 30 | 31.00 | 70 | 66 | 59 | 58 | 63.25 |
| | 1:1 | 1 | 36 | 36 | 35 | 35 | 35.50 | 72 | 85 | 95 | 67 | 79.75 |
| | | 2 | 39 | 36 | 42 | 39 | 39.00 | 97 | 82 | 80 | 80 | 84.75 |
| | | 3 | 37 | 42 | 38 | 40 | 39.25 | 109 | 94 | 90 | 83 | 94.00 |
| | 4:1 | 1 | 48 | 49 | 47 | 47 | 47.75 | 129 | 138 | 107 | 98 | 118.00 |
| | | 2 | 47 | 49 | 48 | 50 | 48.50 | 149 | 144 | 131 | 115 | 134.75 |
| | | 3 | 58 | 53 | 53 | 56 | 55.00 | 164 | 158 | 145 | 127 | 148.50 |
| | 16:1 | 1 | 64 | 84 | 65 | 83 | 74.00 | 219 | 233 | 203 | 169 | 206.00 |
| | | 2 | 63 | 71 | 68 | 65 | 66.75 | 230 | 249 | 238 | 192 | 227.25 |
| | | 3 | 70 | 73 | 70 | 77 | 72.50 | 241 | 263 | 245 | 190 | 234.75 |
| 1 | 1:4 | 1 | 60 | 68 | 70 | 86 | 71.00 | 166 | 175 | 153 | 128 | 155.50 |
| | | 2 | 76 | 79 | 74 | 72 | 75.25 | 175 | 200 | 181 | 159 | 178.75 |
| | | 3 | 96 | 101 | 89 | 79 | 91.25 | 190 | 213 | 202 | 174 | 194.75 |
| | 1:1 | 1 | 73 | 95 | 88 | 88 | 86.00 | 221 | 225 | 206 | 185 | 209.25 |
| | | 2 | 84 | 132 | 99 | 100 | 103.75 | 267 | 285 | 262 | 243 | 264.25 |
| | | 3 | 98 | 118 | 114 | 135 | 116.25 | 273 | 319 | 313 | 275 | 295.00 |
| | 4:1 | 1 | 106 | 124 | 111 | 130 | 117.75 | 301 | 329 | 333 | 282 | 311.25 |
| | | 2 | 122 | 219 | 127 | 130 | 149.50 | 372 | 376 | 378 | 343 | 367.25 |
| | | 3 | 129 | 147 | 139 | 158 | 143.25 | 395 | 381 | 418 | 407 | 400.25 |
| | 16:1 | 1 | 144 | 184 | 158 | 155 | 160.25 | 615 | 513 | 507 | 459 | 574.33 |
| | | 2 | 155 | 199 | 244 | 164 | 190.50 | 538 | 549 | 552 | 533 | 543.00 |
| | | 3 | 170 | 284 | 194 | 176 | 206.00 | 583 | 606 | 586 | 542 | 579.25 |
| 2 | 1:4 | 1 | 230 | 234 | 301 | 219 | 246.00 | 498 | 491 | 518 | 468 | 493.75 |
| | | 2 | 237 | 269 | 284 | 286 | 269.00 | 643 | 717 | 663 | 585 | 652.00 |
| | | 3 | 268 | 316 | 297 | 282 | 290.75 | 727 | 778 | 734 | 620 | 714.75 |
| | 1:1 | 1 | 237 | 271 | 275 | 263 | 261.50 | 581 | 801 | 693 | 627 | 675.50 |
| | | 2 | 349 | 380 | 366 | 353 | 362.00 | 1015 | 1010 | 940 | 849 | 953.50 |
| | | 3 | 366 | 454 | 438 | 418 | 419.00 | 1213 | 1226 | 1098 | 969 | 1126.50 |
| | 4:1 | 1 | 306 | 325 | 346 | 343 | 330.00 | 904 | 907 | 888 | 802 | 875.25 |
| | | 2 | 414 | 462 | 490 | 437 | 450.75 | 1428 | 1408 | 1189 | 1063 | 1272.00 |
| | | 3 | 482 | 504 | 519 | 487 | 498.00 | 1460 | 1639 | 1398 | 1214 | 1427.75 |
| | 16:1 | 1 | 423 | 417 | 435 | 377 | 413.00 | 1467 | 1347 | 1242 | 1111 | 1291.75 |
| | | 2 | 516 | 532 | 527 | 502 | 519.25 | 2049 | 1731 | 1540 | 1342 | 1665.50 |
| | | 3 | 600 | 636 | 579 | 568 | 595.75 | 2373 | 2030 | 1731 | 1556 | 1922.50 |
| 3 | 1:4 | 1 | 844.5 | 920 | 799 | 798 | 843.23 | 2574 | 2330 | 1912 | 1655 | 2117.75 |
| | | 2 | 893 | 1171 | 1048 | 1194 | 1076.50 | 3771 | 3387 | 2693 | 2244 | 3023.75 |
| | | 3 | 1000 | 1270 | 1318 | 1110 | 1174.50 | 4066 | 3881 | 2970 | 2481 | 3349.50 |
| | 1:1 | 1 | 783 | 1106 | 1010 | 932 | 957.75 | 4401 | 3155 | 2750 | 2228 | 3133.50 |
| | | 2 | 1498 | 1533 | 1567 | 1470 | 1517.00 | 6157 | 5217 | 3992 | 3243 | 4652.25 |
| | | 3 | 1749 | 2017 | 1888 | 1705 | 1839.75 | 6818 | 5885 | 4755 | 3793 | 5312.75 |
| | 4:1 | 1 | 1097 | 1339 | 1170 | 1077 | 1170.75 | 5851 | 4490 | 3302 | 2538 | 4045.25 |
| | | 2 | 1751 | 2012 | 1744 | 1632 | 1784.75 | 8029 | 6460 | 4972 | 3733 | 5798.50 |
| | | 3 | 1823 | 2184 | 2162 | 1816 | 1996.25 | 8498 | 7921 | 6246 | 4475 | 6785.00 |
| | 16:1 | 1 | 1216 | 1450 | 1348 | 1229 | 1310.75 | 6815 | 4901 | 3794 | 3274 | 4696.00 |
| | | 2 | 1820 | 2002 | 1976 | 1715 | 1878.25 | 10745 | 8655 | 5332 | 4430 | 7290.50 |
| | | 3 | 1998 | 2465 | 2271 | 2057 | 2197.75 | 12061 | 10248 | 6514 | 5102 | 8481.25 |

## 5. Summary

The paper proposes a new model for representing Cluster-Tools with two load locks and provides generalization of an existing Cluster-Tool model Ortner [21] that was restricted to three chambers. A computational study indicates that on average the generalized model for three chambers is a little faster than the new model proposed in this paper, but for several instances the proposed alternative model seems to have small but significant benefits.

For four chambers, the new model performs significantly better than the generalized model. Except for very small instances, the new model out-performs the generalized one.

Our future research will consider Cluster-tools with more than two load locks and we will investigate the impact of conflicting recipes. We will also work on sensitivity analysis and robustness with respect to recipe combinations and availability.

## AppendixA. Algorithm to find a minimal representation for the dual based model

With respect to (6-9) it is important to keep the set $\mathcal{K}$ as small as possible. Therefore, the concept of redundant vectors (redundant minimal basic cuts) will be introduced and different criteria for redundancy will be stated and proven. According to (38), vectors $\Pi_k := (\Pi_{r,k}) \in \mathbb{R}^{\mathcal{R}}$ are used as a lower bound on $\rho$, i.e. $\langle 1 - \Pi_k, x\rangle \leq \rho$. Suppose for two vectors $\Pi_k$ , $\Pi_{k'}$ the following condition true: $\langle 1 - \Pi_k, x\rangle \leq \langle 1 - \Pi_{k'}, x\rangle \leq \rho$. Then $\Pi_{k'}$ dominates $\Pi_k$ and $\Pi_k$ can be removed. The condition can be reformulated to $\langle \Pi, x\rangle \geq \langle \Pi_{k'}, x\rangle$ and leads to the following more general definition of redundancy:

**Definition 2.** Assuming that a set $A$ of $m$ vectors, $A = \{a_1, a_2 \ldots, a_m\}$ with $a_i \in \mathbb{R}^n$ and $a_i \geq 0$ is given. Then a vector $b$ is called *redundant* if the following is true:

$\forall x \geq 0 : \langle b, x\rangle \geq \min_i \langle a_i, x\rangle$

Or, in other words $b$ is called redundant if $\forall x \geq 0 \exists i : \langle b, x\rangle \geq \langle a_i, x\rangle$. Note, that it is sufficient to consider vectors $x$ of unit length, therefore $b$ is called redundant if:

$\forall x \geq 0, \|x\| = 1 \exists i : \langle b, x\rangle \geq \langle a_i, x\rangle \Leftrightarrow$

$\forall x \geq 0, \|x\| = 1 \exists i : \langle 1 - b, x\rangle \leq \langle 1 - a_i, x\rangle \Leftrightarrow$

$\forall x \geq 0, \|x\| = 1 \exists i : \|\langle 1 - b, x\rangle x\| \leq \|\langle 1 - a_i, x\rangle x\|$

Geometrically, $\langle 1 - b, x\rangle x$ can be interpreted as the projection of $1 - b$ on $x$. Also note that

$\left\|\langle 1 - b, x\rangle x - \frac{1-b}{2}\right\| = \left\|\frac{1-b}{2}\right\|$ and $\left\|\langle 1 - a_i, x\rangle x - \frac{1-a_i}{2}\right\| = \left\|\frac{1-a_i}{2}\right\|$ for $\|x\| = 1$, therefore the statement is equivalent to:

$U^+_{\|\frac{1-b}{2}\|}(\frac{1-b}{2}) \subset \bigcup_i U^+_{\|\frac{1-a_i}{2}\|}(\frac{1-a_i}{2})$

where $U^+_\epsilon(b) = \{x \in \mathbb{R}^n : x \geq 0, \|x - b\| \leq \epsilon\}$.

Figure A.8 depicts an example in three dimensions. If the sphere $U^+_{\|\frac{1-b}{2}\|}(\frac{1-b}{2})$ is contained in $\bigcup_i U^+_{\|\frac{1-a_i}{2}\|}(\frac{1-a_i}{2})$ then $b$ is redundant. The geometric interpretation can be used to compute if a vector $b$ is redundant or not:

If $b$ is not redundant, then some part of the boundary of $U^+_{\|\frac{1-b}{2}\|}(\frac{1-b}{2})$ defines the boundary of the union of spheres $\bigcup_i U^+_{\|\frac{1-a_i}{2}\|}(\frac{1-a_i}{2})$. The boundary of the union of spheres $\bigcup_i U^+_{\|\frac{1-a_i}{2}\|}(\frac{1-a_i}{2})$ can be described by transformed facets like in Figure A.8 where the white region is touching three neighboring regions. If it is possible to find two facets that belong to two spheres $(i, j)$ that are neighboring to the sphere $U^+_{\|\frac{1-b}{2}\|}(\frac{1-b}{2})$, then a hyperplane $H_{ij}$ exists that contains the intersection of the boundaries of the spheres (in 3d it contains a circle). Furthermore, if $b$ is not redundant this hyperplane has a nonempty intersection with the facet that corresponds to $b$. Therefore, a vector can be found that is on the hyperplane and outside the union. If no such two spheres can be found then the dominating part of the region intersects with at least one of the hyperplanes that are defined by $x_i = 0$.

Therefore, $b$ is not redundant if:

$U^+_{\|\frac{1-b}{2}\|}(\frac{1-b}{2})|_{H_{ij}} \not\subset \bigcup_i U^+_{\|\frac{1-a_i}{2}\|}(\frac{1-a_i}{2})|_{H_{ij}}$ or

$U^+_{\|\frac{1-b}{2}\|}(\frac{1-b}{2})|_{x_i=0} \not\subset \bigcup_i U^+_{\|\frac{1-a_i}{2}\|}(\frac{1-a_i}{2})|_{x_i=0}$

The dimension of the corresponding sub problems is $n - 1$, the method can be reapplied until the dimension of all subproblems is one. Note that if $b$ is not redundant then it is possible to identify one or more spheres $U^+_{\|\frac{1-a_i}{2}\|}(\frac{1-a_i}{2})$ that contain b, the set of the corresponding indices is called $M$, and $M \neq \emptyset$, otherwise it is trivial that $b$ is not redundant. Also suppose that $U^+_{\|\frac{1-b}{2}\|}(\frac{1-b}{2})$ is not fully contained in one of the $U^+_{\|\frac{1-a_i}{2}\|}(\frac{1-a_i}{2})$ $(i \in M)$, otherwise $b$ is clearly dominated. For the procedure it is not necessary to consider all possible $H_{ij}$, it is sufficient to assume that $i \in M$. The corresponding procedure is not very effective and the worst case time complexity is larger than $\mathcal{O}(m^n)$.

In the following, an alternative method, that is based on convex hulls is presented:

**Proposition 3.** *criterion for redundancy: A vector b is redundant with respect to a set of vectors $\{a_i\}$, if and only if it is contained in the convex hull $\langle\{a_i\}\rangle$.*

*Proof.* the negation redundancy is the following:

$$\exists x \geq 0 \quad \forall i: \quad \langle b, x\rangle > \langle a_i, x\rangle$$

$$\exists x \geq 0 \quad \forall i: \quad \langle b - a_i, x\rangle > 0 \tag{A.1}$$

According to Gale [7] (Theorem 2.10, p. 49, which is a corollary of Farkas' Lemma) the corresponding elimination criterion has the form:

$$\exists \lambda \geq 0(\lambda \neq 0) \quad \textstyle\sum_{i\neq j} \lambda_i(b - a_i) \leq 0$$

It is also possible to assume that $\sum_{i\neq j} \lambda_i = 1$ and therefore:

$$\exists \lambda_i > 0 \quad b \leq \textstyle\sum_{i\neq j} \lambda_i a_i$$

This condition states that b is contained in the convex hull of $\{a_i\}_{i\neq j}$. Therefore the problem is solvable by calculating the convex hull; and according to Chazelle [2], the convex hull is computable in $\mathcal{O}(m \log(m) + m^{\frac{n}{2}})$ time. □

A further alternative that avoids the calculation of the convex hull ( see Chazelle [2]) is presented in the following proposition. Among the presented methods it is the most efficient one.

**Proposition 4.** *criterion for redundancy: A vector b is redundant if and only if the following LP is infeasible.*

$$\begin{aligned} \min \qquad & 1^t x \\ & (b - a_i)^t x \geq 1 \quad \forall i \\ & x \geq 0 \end{aligned} \tag{A.2}$$

*Proof.* ⇒: A feasible solution $x$ of the LP also satisfies (A.1) and therefore the vector $b$ is redundant - the objective is irrelevant.

⇒: On the other hand, $b$ is redundant and satisfies (A.1), therefore a vector $x'$ and a positive number $\epsilon > 0$ exist such that $(b - a_i)^t x > \epsilon$. Hence, $\frac{x}{\epsilon}$ is a feasible solution of the LP and the proof is complete. The complexity is defined by the complexity of linear programming algorithms. Algorithms with $\mathcal{O}(n^3 L)$ are known, where $L$ is the bit length of the input data. □

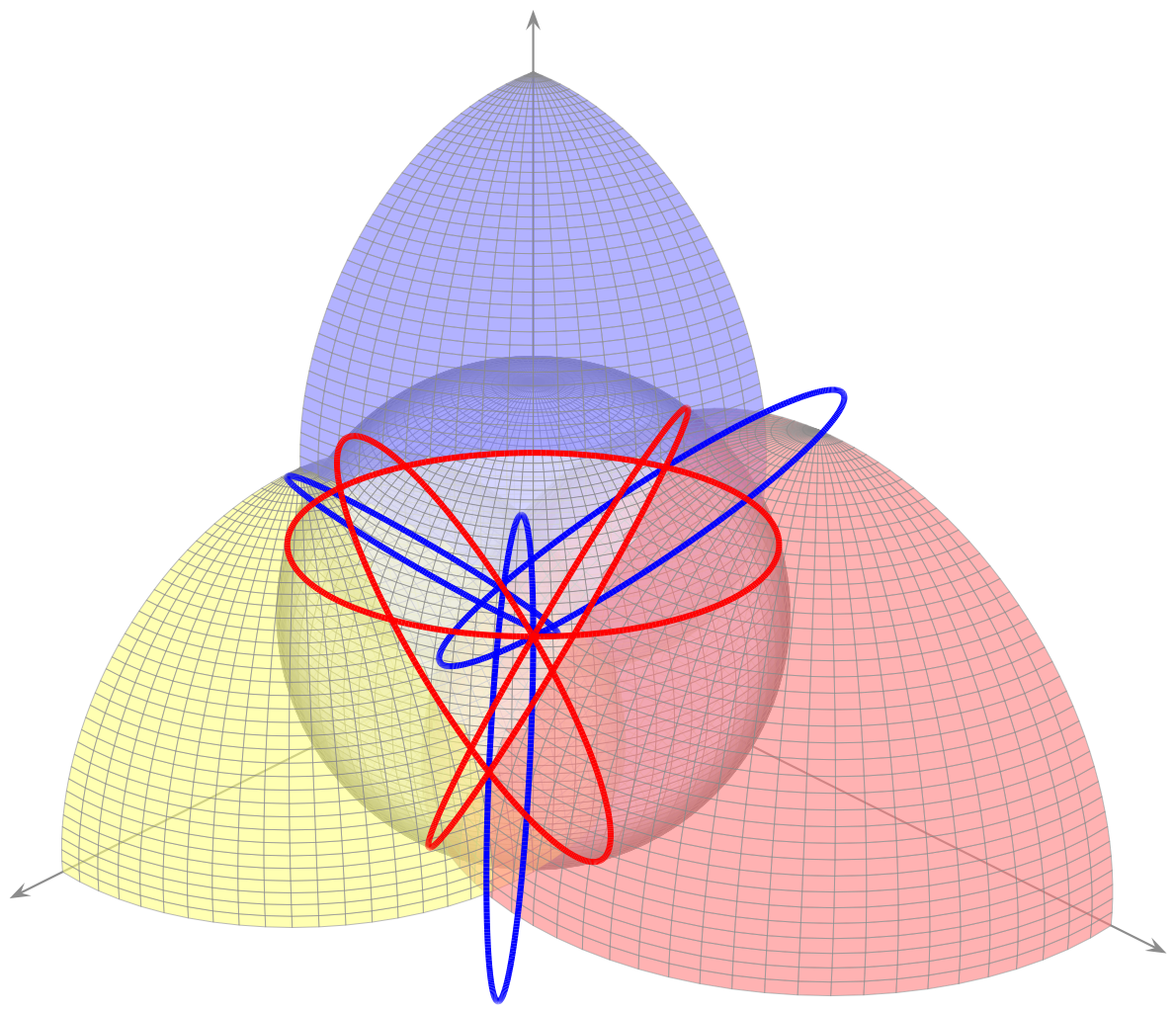

Figure A.8: A three dimensional example with four vectors, where the corresponding set of vectors that is non-dominating is represented by the union of spheres.

| | *cut* | $A$ | $B$ | $C$ | $AB$ | $AC$ | $BC$ |
|---|---|---|---|---|---|---|---|
| $\Pi_1$ : | $\{A, B, AB, \widetilde{A}, \widetilde{B}, \widetilde{AB}\}$ | 1 | 1 | | 1 | | |
| $\Pi_2$ : | $\{A, B, C, \widetilde{A}, \widetilde{B}, \widetilde{C}\}$ | 1 | 1 | 1 | | | |
| $\Pi_3$ : | $\{A, B, C, AB, \widetilde{A}, \widetilde{B}\}$ | 1 | 1 | $\frac{1}{2}$ | $\frac{1}{2}$ | | |
| $\frac{1}{2}(\Pi_1 + \Pi_2)$ : | | 1 | 1 | $\frac{1}{2}$ | $\frac{1}{2}$ | | |

Table A.14: minimal basic cuts and redundancy

**Definition 5.** Note, that a set of vectors $A = \{a_i \in [0, \infty)^n\}$ is called *minimal* if none of its members is redundant with respect to $A \setminus \{a_i\}$, more precisely:
$\forall j \exists x \geq 0 \quad \forall i \neq j : \quad {a_j}^t x < a_i^t x.$

Finally, to show that not all minimal basic cuts are needed in the LP formulation, a simple example that considers three chambers is presented:

**Example 6.** For three chamber, Table A.14 lists three of the minimal basic cuts, named $\Pi_1, \Pi_2, \Pi_3$. Obviously, $\Pi_3 = \frac{1}{2}\Pi_1 + \frac{1}{2}\Pi_2$, therefore the criterion formulated in Proposition 3 shows that $\Pi_3$ is redundant with respect to $\{\Pi_1, \Pi_2\}$.